# The RED-100 experiment

**D.Yu. Akimov,**[a*] **I.S. Alexandrov,**[a,b] **R.R. Alyev,**[c] **V.A. Belov,**[a,d] **A.I. Bolozdynya,**[a]
**A.V. Etenko,**[a] **A.V. Galavanov,**[a,e] **E.M. Glagovsky,**[a] **Y.V. Gusakov,**[a,e] **A.V. Khromov,**[a,b]
**S.M. Kiselev,**[c] **A.M. Konovalov,**[a] **V.N. Kornoukhov,**[a,f] **A.G. Kovalenko,**[a,d]
**E.S. Kozlova,**[a,d] **A.V. Kumpan,**[a,b] **A.V. Lukyashin,**[a] **A.V. Pinchuk,**[a] **O.E. Razuvaeva,**[a,d]
**D.G. Rudik,**[a] **A.V. Shakirov,**[a] **G.E. Simakov,**[a,d] **V.V. Sosnovtsev**[a] **and A.A. Vasin**[a]

[a] *National Research Nuclear University "MEPhI" (Moscow Engineering Physics Institute),*
 *Moscow, 115409, Russian Federation*

[b] *National Research Tomsk Polytechnic University,*
 *Tomsk, 634050, Russian Federation*

[c] *Branch of "Rosenergoatom" Concern JSC "Kalinin Nuclear Power Plant"*
 *Udomlya, Tver region, 171841, Russian Federation*

[d] *National Research Center "Kurchatov Institute",*
 *Moscow, 123182, Russian Federation*

[e] *Joint Institute for Nuclear Research, Dubna,*
 *Moscow region 141980, Russian Federation*

[f] *Institute for Nuclear Research,*
 *Moscow, 117312, Russian Federation*

 *E-mail:* `DYAkimov@mephi.ru`

ABSTRACT: The RED-100 two-phase xenon emission detector has been deployed at 19-m distance from the reactor core of the Kalinin Nuclear Power Plant (KNPP) in 2021 – 2022 for investigation of the possibility to observe reactor antineutrinos using the effect of coherent elastic neutrino-nucleus scattering (CEvNS). The performance of the main systems of the RED-100 setup at operating nuclear power plant is described. There is no correlation of the radioactive background at the experimental setup site with ON and OFF states of the reactor. The data taking run was carried out at the beginning of the year 2022 and covered both the reactor OFF and ON periods.

KEYWORDS: Noble liquid detectors (scintillation, ionization, double-phase); Neutrino detectors; Very low-energy charged particle detectors

---

[*] Corresponding author

# Contents

## 1. Introduction

The possibility of using neutrino detectors to improve safety of nuclear energy and for nonproliferation of nuclear weapons monitoring has been studied for more than 40 years [1]. The neutrino method for monitoring nuclear power reactors was first proposed and experimentally studied by the scientists from the National Research Center "Kurchatov Institute" in experiments at the Rovno NPP in the 1980s. Data obtained in these experiments showed the principal possibility of the real-time independent monitoring of the process of nuclear fuel burnup, and evaluation of the plutonium content in the fuel [2], [3]. By the early 2000s, IAEA experts have recognized the usefulness of the neutrino method for remote monitoring of the reactor core and supported its practical use for the above mentioned purposes [4], [5]. Implementation of the neutrino monitoring method into the real operating procedures of a nuclear power plant or other nuclear industrial facilities must meet the following specific requirements:

    - detection should be remote and carried out at a distance of at least 10-20 m from the reactor core outside the biological shield;

    - detecting setup must be compact enough to be placed in the existing technological rooms of NPP.

The main technical problem in the implementation of neutrino monitoring of a nuclear reactor core is the extremely small cross-section of the neutrino interaction with matter. Until recently, the reaction of inverse beta decay was almost exclusively used to detect reactor neutrinos. With the discovery in 2017 of the coherent elastic neutrino-nucleus scattering (CEvNS) [6] which has a cross-section several orders of magnitude larger than the cross-section of inverse beta decay an attractive prospect arose of using this process to build neutrino detectors of a new type that meet the necessary requirements. The main challenge of reactor neutrino detection by the CEvNS process on heavy nuclei for which the interaction cross-section is the largest is an extremely weak ionization signal (down to single electrons). This signal must be reliably detected in a fairly massive (hundreds of kilograms) detector. To solve this technical problem, the RED-100 collaboration proposed to use the technology of two-phase emission detectors [7].

A two-phase emission detection technique proposed at MEPhI for the first time about 50 years ago [8] provides a unique possibility to build large-mass detectors based on condensed noble gases and have a superior sensitivity to a small electric charge due to the electroluminescent amplification [9], [10]. Another unique feature of a two-phase detector is the possibility to build a so-called "wall-less detector" [11], when the inner volume of the detector is used for observation of the events, while the outer layer serves as an active veto and shield for external radiation.

The first observation of CEvNS has caused an outbreak of theoretical activity in particle physics that gave new restrictions on physics beyond the Standard Model. The CEvNS process has important implications not only for high energy physics, but also for astrophysics, nuclear physics, and other fields. This motivated development of detectors of large size or based on new technologies in order to increase the sensitivity of experiments and to cover the regime of full quantum-mechanical coherency at lower energies.

At present, RED-100 at KNPP is the most massive detector among all those operating at reactor sites [12], [13], [14], [15], [16], [17], [18], [19], [20] and the only liquid xenon (LXe) detector. The purpose of the experiment is to test the technology of a two-phase emission detector under real operating conditions at a nuclear power plant and to observe the CEvNS process for reactor antineutrinos for the first time. As in all other experiments on the detection of reactor antineutrinos, a two-phase emission detector based on liquid noble gas possesses an internal noise in the energy range of detected CEvNS events (see [21] and references therein). This noise correlates with the total energy deposited in the detector per unit of time and is produced mainly by cosmic muons. Despite the fact that the muon flux at the location of the detector at the power unit is significantly suppressed by the structural elements of the reactor building, the reactor itself and spent fuel pools, this type of background is dominant in the experiment. One of the main technical problems to be solved in the course of work with the RED-100 detector was the development of a trigger mode for recording of useful events, candidates for CEvNS, under conditions of this noise.

## 2. The RED-100 experimental setup at KNPP

The RED-100 experimental setup has been described in detail in our previous publications [21], [22], [23], [24]. All general technical solutions of the RED-100 experimental setup at KNPP were kept practically the same as in our previous test runs in the MEPhI laboratory. For transportation to KNPP the setup was disassembled into parts as small as possible and assembled again in a hall under the 3-GW thermal power reactor of the 4-th KNPP unit.

The RED-100 detector in a copper and water shield having a total mass of ~ 15 t was placed in the technological hall under the reactor at ~ 19-m distance from the reactor core center and ~ 1 m apart from the reactor vertical axis. An updated block diagram of the RED-100 experimental setup is shown in Figure 1 (on top) together with the photo of the setup at KNPP (on bottom).

The setup was being assembled during the period from the end of January 2021 till the middle of April 2021. The first (technical) run continued from the end of April 2021 till the beginning of June 2021. The second run started at the beginning of December 2021 and finished at the beginning of March 2022. In the second run, measurement aimed at search for CEvNS was carried out (from the mid of February 2022), with roughly equal OFF and ON periods of the reactor operation mode.

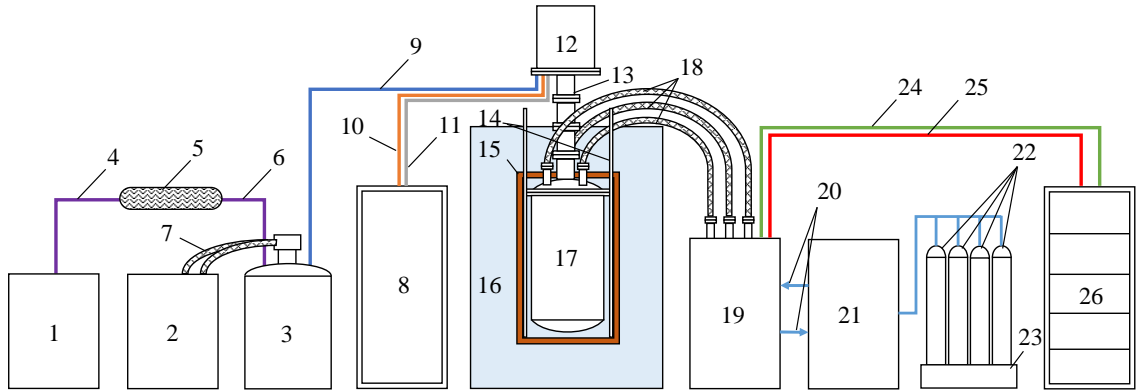

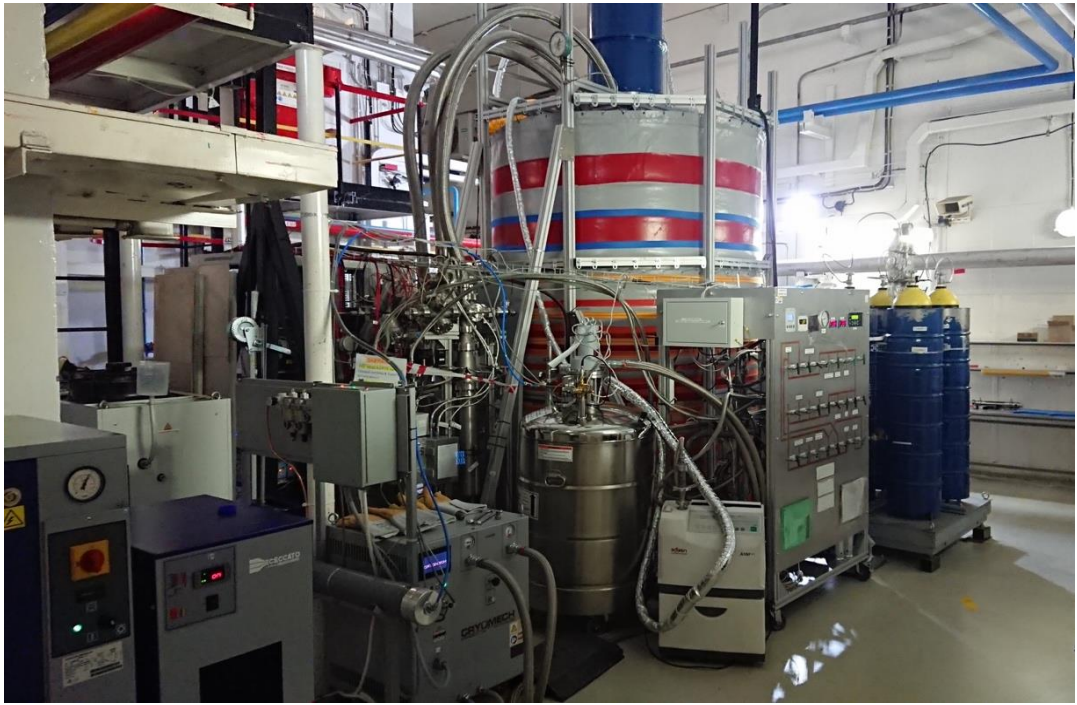

**Figure 1.** Block diagram (on top) and general view (on bottom) of the RED-100 setup at KNPP: 1 – air compressor, 2 – He compressor, 3 – liquid nitrogen Dewar with a cold head, 4 – compressed air line, 5 – nitrogen-oxygen membrane separator, 6 – pure nitrogen line, 7 – high pressure He lines, 8 – thermosyphon control rack, 9 – $LN_2$ cold line, 10 – thermocontrol cable line, 11 – thermosyphon nitrogen supply pipe lines, 12 – cryostat of the thermosyphon cryogenic system, 13 – bellows with thermosyphon pipes, 14 – calibration pipes, 15 – copper shield, 16 – water tank, 17 – RED-100 detector, 18 – bellow conduits, 19 – interface unit, 20 – inlet and outlet Xe lines, 21 – gas purification system with control panel, 22 – xenon storage system, 23 – scales, 24 – signal cabling, 25 – high voltage cabling, 26 – electronics and data acquisition system rack. In the photo, units 8 and 26 are in the background and not seen.

## 3. Subsystems of the RED-100 and their operation

### 3.1 Cryogenics and LN$_2$ supply

Cooling and thermostabilization of the LXe volume in the RED-100 detector is performed by a thermosyphon technology. Principle of thermosyphon operation and performances of this system were described in our previous publication [21]. There are four thermosyphons in the RED-100 cryogenic system. Maintaining of the operation temperatures (–105º ± 0.3º C) of the RED-100 detector is performed by varying thermosyphon conductivities (cooling power) which, in turn, is done by regulating a pressure of gaseous nitrogen inside thermosyphons. This can be done either manually or automatically by filling a thermosyphon pipe with nitrogen from a high-pressure cylinder or by releasing a gas to the atmosphere. The thermosyphon gas system which controls parameters is placed in a separate rack (8 in Figure 1). This system is connected to the RED-100 thermosyphons by armed plastic pipes (11).

All liquid nitrogen required for cooling of the RED-100 detector is produced locally by a LN$_2$ plant LNP-120 [25] by Cryomech, Inc. (1 – 7 in Figure 1). The liquid nitrogen is transferred periodically (approximately, every 15 h) to the cryostat of the thermosyphon cryogenic system through a thermoisolated pipeline (9). Consumption of LN$_2$ by RED-100 is ~ 2 l/h which is significantly lower than a nominal LNP-120 production rate (5 l/h).

### 3.2 Gas system and purification

The purification procedure of our initially very dirty xenon was described in detail in previous publications [21], [26], [27]. The very high degree of xenon purity correspondent to the electron lifetime in LXe of several ms in the RED-100 detector was achieved after spark-discharge purification in a liquid phase followed by long-term continuous circulation in a gas phase through a hot getter (SAES MonoTorr).

The xenon gas system of RED-100 includes four 50-l high-pressure aluminum cylinders (22) situated on a weighting platform of scales METTLER TOLEDO KD1500 (23), a gas purification system with control panel (21), and an interface unit (19). These parts are interconnected by bellow metal hoses. The interface module is connected with the RED-100 detector by three metal bellow conduits (18) which contain the inlet and outlet gas lines (20), high-voltage (24) and signal (25) lines.

A circulation loop includes the following elements: detector, circulation membrane pump (KNF N143 SV.12E), SAES MonoTorr getters, flowmeters/controllers, and gas lines. The main xenon flow path comes through the liquid phase of xenon: the purified gas enters the detector from top, then condenses to the liquid, and the liquid xenon comes out from the bottom of the detector vessel, evaporates in a heat exchanger and goes back to the purification system. In addition, the conduits are constantly purged in the backward direction (towards the interface unit) during the circulation of the Xe gas as they contain a lot of materials which may produce contaminations. For example, the HV cables have many layers of electroisolating Teflon tape. The portion of the gas coming this way is set programmingly by flow controllers.

The values of the electron lifetime in LXe achieved in the first and second runs were ~ 2 and ~ 1 ms, respectively, that was enough for detection of the CEvNS signals with high efficiency, since they are significantly larger than the full drift time of ~ 260 μs.

### 3.3 Slow control

Monitoring and control of the RED-100 setup parameters is performed with the use of a system built on the EPICS platform [28]. Data taking and processing is carried out by four specialized industrially produced controllers. The system is distributed and remains partially functional even when only part of the controllers is in operation.

The slow control system stores the RED-100 setup parameters in a database InfluxDB [29] for later analysis. Information from the database is available with CS-Studio software [30] and/or through Grafana web interface [31].

An operator workplace station serves to visualize the current data and change the RED-100 setup parameters, to plot time-dependences of various parameters in order to analyze long-term trends, and to control the DAQ system. There are two functionally identical operator stations. One of them is located outside the reactor building, and the other station is placed nearby the setup. The second one serves to control the setup when the presence of an operator at the experimental site is necessary: during filling and evacuation of xenon, filling the thermosyphon Dewar by $LN_2$, calibration by gamma-sources, etc.

### 3.4 Electronics and DAQ

The composition and characteristics of the electronics used in the RED-100 setup are governed by the parameters of signals produced in the detector of this type. Regular events with a large deposited energy, from several tens of keV and higher, consist of a prompt short scintillation signal (S1) and a delayed much bigger electroluminescent signal (S2). The characteristic duration of S1 is several tens of nanoseconds, while that of electroluminescence is about 2 μs. The maximum interval between S1 and S2 is equal to the full drift time, i.e. ~ 260 μs. For extremely low-energy events having several ionization electrons which are of our interest for CEvNS signals, S1 is missed, and events usually consist of S2 only. In this case, S2 signals are represented by individual single photoelectrons (SPE) signals from photomultipliers.

All signals in the active region of RED-100 are detected by two arrays of PMTs (Fig. 2). The PMTs used in RED-100 are Hamamatsu R11410-20 VUV-sensitive metal packed low-background devices designed for operation at cryogenic temperatures. Since a metal body of PMT has electrical connection to a photocathode, a positive voltage supply is used to power PMTs. The PMTs of this type provide a high gain that results in the amplitude of the SPE pulses of an order of 1 mV. In the bottom array, only seven central PMTs (see Figure 2) are used: four with nominal voltage for triggering on scintillation, and the rest three, with lowered voltage, for muon veto (see below). Each PMT is connected to a vacuum multipin feedthrough on the interface unit with a single ~ 5-m coaxial cable used both for signal readout and HV supply. All these cables are placed inside one of the flexible bellow conduits (Fig.1, 18). A decoupling box is connected to each feedthrough and suits six channels. In the box, each channel includes a decoupling capacitor and a passive protection circuit from possible HV breakdown. The box is connected by 7-m coaxial cables to the electronics rack (Fig.1, 24). The PMT signals are amplified by Phillips Scientific 777 fast low-noise 8-channel amplifiers with a bandwidth of 0 – 200 MHz and adjustable gain tuned to ~10. Each amplifier has two outputs, one connected to ADC, and the other, to Phillips Scientific 748 Line Splitter, which is used to split the signals for the trigger circuits. Flash ADC CAEN V1730 units (500 MHz sampling rate, 14 bits resolution) are used for signal digitization. The high voltages of the PMTs are adjusted to equalize the gains of the individual PMTs, so the mean amplitude of the SPE signals in DAQ is equal to 8 mV.

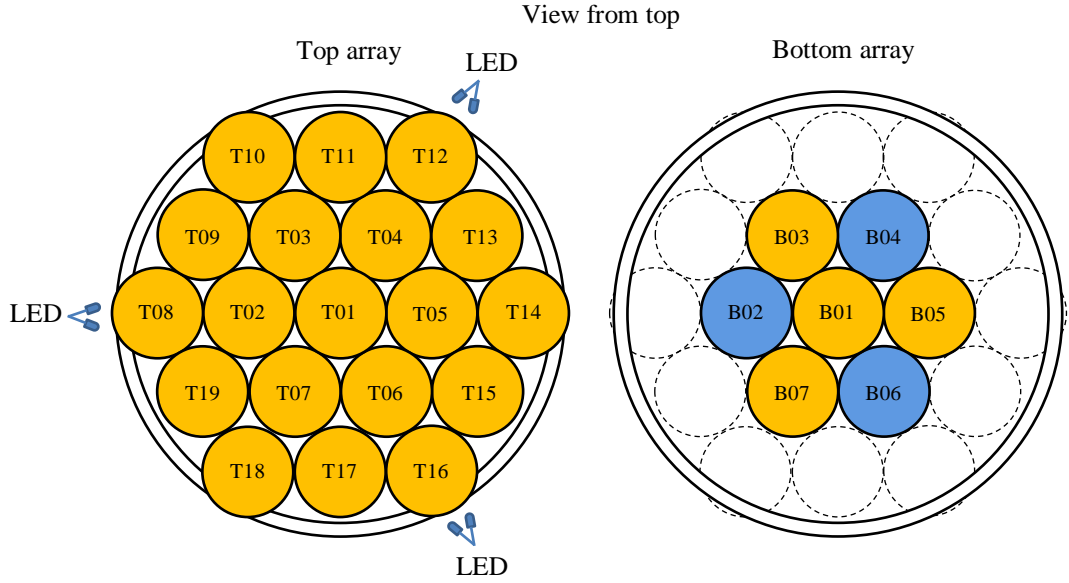

**Figure 2.** The map of active channels in two PMT arrays of the RED-100 detector during its operation at KNPP; a projection of electrode ring-holders is shown in both views; the PMTs used for producing a veto pulse are shown by blue.

Signal waveforms are recorded for channels T1 – T19 and B1, B3, B5, B7 with 300-μs and 30-μs full duration of a sampling window. The former is used for events in which we want to observe a full drift time: gammas (from the background or the calibration sources), muons, LED-calibration and random_SE events (see below for the last two types). The latter is used to record the CEvNS-like events of interest.

The electronics contains also a custom unit which serves for applying a pulse (350 V, 0.3 – 6 ms) to the electron shutter (see description in [21]). This shutter prevents PMTs from overloading and aging by the enormously large muon signals and helps to reduce a spontaneous single electron (SE) emission. The reverse (positive) polarity is applied to the bottom electrode of the shutter through a decoupling capacitor wherein both electrodes are fed from the same HV supply through independent resistive-capacitive filters.

The whole electronics rack is equipped with an air-condition system for cooling and maintaining the necessary temperature inside the rack. Temperature stabilization of the electronics is crucial because the ambient temperature in the detector location varies from < 20°C to 40°C.

## 3.5 Trigger

A trigger system of the RED-100 detector is required to catch events in a very large energy range from sub-keV region to MeV-scale and even to hundred MeV for cosmic muons. In order to fulfil this challenging task, the core trigger logic was built using CAEN V1495 customizable FPGA unit. The trigger system has several major modes of operation to cover all required event types. The trigger mode is set by DAQ software depending on a measurement task.

The CEvNS signals of interest are expected to consist of up to 6 ionization electrons (see [21]). Such small signals are represented by a bunch of the SPE pulses coming from different PMTs in the array and distributed nearly uniformly in a 2-µs interval (see an example of such an event in Figure 3). In order to achieve good trigger parameters for these signals, a dedicated logic was created [32]. This logic is based on counting SPE pulses from the top PMT array within a 2-µs time window. The pulses for it come from the CAEN V895 leading edge discriminators with a threshold corresponding to ~60% acceptance efficiency of an SPE pulse for each PMT channel. The trigger is generated if a total number of SPEs lies within the specified limits. The low threshold number corresponds to 2 – 3 SE depending on the event position. Additional veto is provided for this trigger mode in order to remove time periods when there is a noticeable background light load inside the detector due to previous large muon signals. This veto is also based on SPE counting, and it is generated if there were more than 50 SPEs during the 50-µs interval.

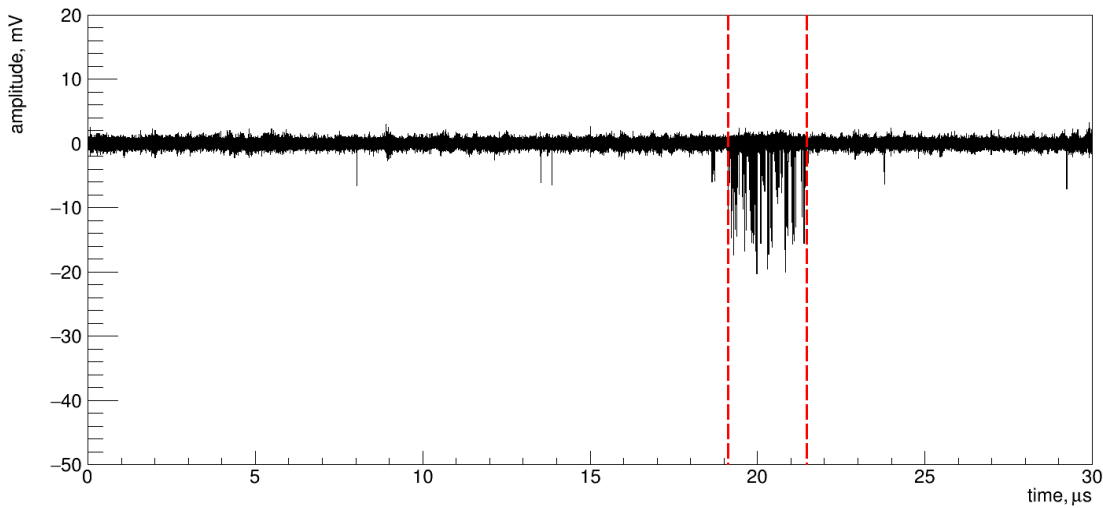

**Figure. 3.** Waveform of the CEvNS candidate event with a magnitude of ~ 3 SE.

The gamma and muon modes operate with a threshold trigger on a scintillation. For this purpose, the signals from B1, B3, B5, and B7 channels (see Figure 2) are summed by the CAEN N625 linear fan-in-fan-out unit. The trigger is based on a leading-edge discriminator with time-over-threshold feature (Phillips Scientific 711) with a threshold of 100 mV and additional selection for pulses shorter than 250 ns to trigger on S1 only. It also provides a gamma veto with a fixed duration of 100 µs for low-energy trigger modes. In the muon mode, a leading-edge 100-mV threshold discriminator is used, however, the PMT HV is reduced to provide the gain equal to ~1:10-th from the nominal gain in other modes.

There are two additional service modes: LED-calibration and random_SE. Both modes are triggered from a pulse generator with a frequency of ~2 Hz. The former is used to calibrate and to monitor SPE peaks of individual PMTs. The LEDs operating at low-intensity in a pulse mode synchronized with the pulse generator seed the whole 300-µs waveform with randomly distributed individual SPEs. The random_SE mode is intended to study SPE and SE signals using spontaneous SE emission events. The big advantage of this mode is that it provides effectively zero threshold for SE detection.

The trigger logic unit generates a muon veto pulse for the electron shutter. The muon is detected using signals from B1, B3, B5, and B7 channels with a high threshold (1V) discriminators and majority of 2. It is known that SE noise after muon depends on a deposited charge and decreases with time, hence a variable electron shutter pulse duration is used. The shutter pulse duration is derived from S1 pulse duration, which for a given threshold, depends on a deposited charge. For this purpose, the sum signal from channels B2, B4, and B6 (see Figure 2) produced by the CAEN N625 unit is used. These PMTs are powered with the lower HV, and thus, operate at a reduced gain and keep linear response even for muon signals. The sum signal goes to a leading-edge discriminator with a time-over-threshold feature and 100-mV threshold. A relative magnitude of S1 is obtained by measuring a discriminator pulse width. The width varies from 0 to 400 ns, which provides 16 possible values given 25-ns FPGA clock. The shutter pulse is produced from this width using look-up table and results with a duration from 0.3 to 6 ms depending on the value of S1.

There is an additional technical veto for 300 μs after DAQ trigger to prevent repetitive triggering during the event waveform digitizing and recording by DAQ. Due to hardware limitations, a so-called "decimation" of triggers is used when the trigger rate substantially exceeds the ultimate recording rate of the DAQ system. This is done by a random coincidence of a trigger and a generator pulse with 1 kHz period and variable width. In this case the reduction factor of the real trigger rate is equal to the duty factor of the generator pulse. Typical values of this factor are shown in Table 1.

In order to properly account for intervals of trigger veto of all types and carefully calculate live time for reliable rate measurement, a dedicated logic is used. During a data taking run, pulses from a 1-MHz generator in coincidence with veto sources enabled for the used mode are continuously accumulated, thus providing the exact value of live time.

**Table 1.** Typical values of duty factor for different trigger modes.

| Trigger mode | CEvNS | gamma (with water shield) | gamma (without water shield) | muon |
|---|---|---|---|---|
| Duty factor | 1 | 0.06 | 0.004 | 0.33 |

### 3.6 Radiation shield and gamma-calibration tubing

A radiation shield of the RED-100 detector described in detail in [24] is shown in Figure 4 together with a cross-section of the RED-100 detector. It includes ~ 70 cm of water and 5 cm of copper. Four plastic pipes (5 in Figure 4) are installed inside the copper shield along the corners for calibration by a gamma-source ($^{137}$Cs and $^{60}$Co) suspended on a wire.

### 3.7 Ambient radiation measurements

Radiation background outside the water tank is measured and monitored during the data-taking run by means of the following detectors: scintillation gamma spectrometer NaI[Tl] (diam. 80 x 80 mm), neutron liquid scintillator detector Bicron (model 3MAB-1F5BC501A; diam. 3 x 5"), and two domestic radon indicators RADEX, MR107 and MR107+ models. Also, before the assembling of the RED-100 setup, the gamma background was measured in various points of the experimental hall by the gamma spectrometer NaI[Tl] of bigger sizes (diam 150 x 100 mm). The NaI[Tl] and Bicron BC501A detectors are placed at two different scaffoldings approximately at a height of 1.8 m from the floor.

Signals from the NaI[Tl] detector are amplified and shaped with ORTEC 572A NIM unit, then digitized with ORTEC MCA, and stored with MAESTRO Software.

Signals from the Bicron detector are split into two identical independent fast amplifiers (CAEN N978) with different by a factor of 4 gains (high- and low-sensitivity channels). The signals are digitized with 500 MS/s frequency by a LeCroy WaveRunner 640Zi digital oscilloscope and stored to its disk in an event by event mode for offline analysis. The signal from the amplifier output of the high-sensitivity channel is sent to a leading-edge threshold discriminator to produce a trigger to the oscilloscope. A logical pulse scaler is used to count these triggers in order to obtain a real detector count rate since a digital oscilloscope has uncontrollable "dead time". For neutron calibration (to obtain a neutron median on a PSD plot) the Bicron detector was brought into an adjoining room with a reactor monitoring equipment containing a PuBe source.

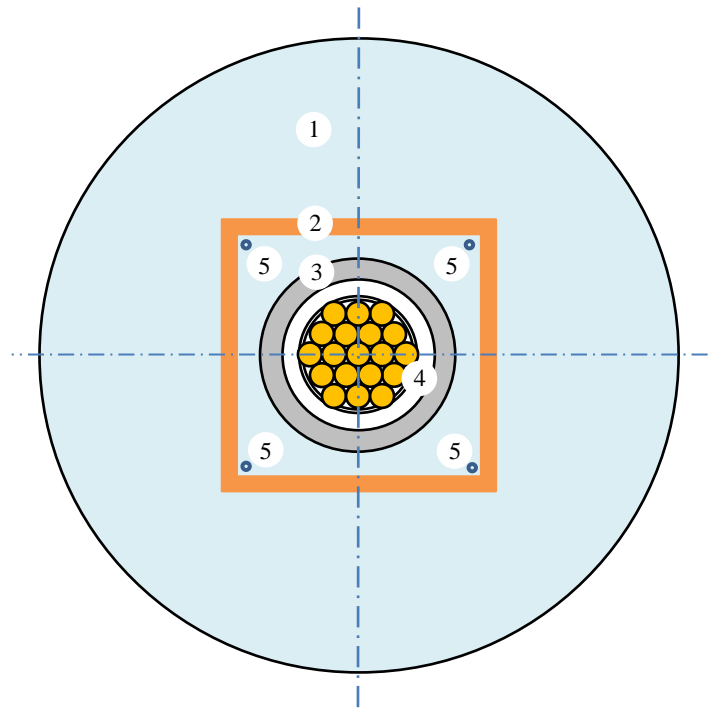

**Figure 4.** Schematic layout of the RED-100 detector elements inside the passive shield;
1 – water tank (ID 2.2 m), 2 – copper castle (size 0.9 m, thickness 5 cm), 3 – vacuum vessel (ID 0.64 m), 4 – cold vessel (ID 0.5 m, filled by ~ 6-cm Teflon), 5 – plastic pipes (OD 2 cm); PMTs are shown by orange circles.

The RADEX detectors are installed at ~ 1 m distance from the water tank. They have a sensitivity threshold of 30 Bq/m$^3$. The data from the RADEX detectors are recorded internally and then downloaded to a laptop.

### 3.8 Cosmic muons measurement

Measurement of the muon signals with the RED-100 detector is very important for understanding of both the background conditions and detector performance. Muons are recorded both by DAQ (with full-channel information for each event) and by the LeCroy WaveRunner 640Zi digital oscilloscope (aggregated).

The main purpose of the muon signal measurement by the digital oscilloscope is to obtain the electron lifetime in LXe. For this purpose, the summed analogue PMT signals (from the top and bottom PMT arrays separately) are averaged by oscilloscope's internal data analysis code. A threshold trigger with additional condition of having a pulse with < 450 ns duration (scintillation) at the beginning of a waveform is used. The total number of events collected in each trigger mode sample is ~ 10000. Then the averaged waveforms are recorded, corrected for a ballistic deficit and approximated by an exponent to obtain the electron lifetime value.

## 4. Monitoring of the important parameters

### 4.1 General detector parameters

The main important parameters of the detector such as detector pressure, temperature of the LXe cold volume, circulation rate, residual pressure between the inner and outer vessels of the detector cryostat, and temperature of the electronics were continuously recorded to a slow-control database and monitored by an operator throughout the setup operation.

The pressure variation caused by the LXe temperature instability and the instability of the circulation rate was from 1.25 to 1.3 bar during the data taking period. This corresponds to ~ 5% variation in the EL gain that is not significant taking into account much worse energy resolution in the keV and sub-keV energy region.

The residual pressure between the inner and the outer vessel of the detector cryostat was maintained at a level of ~ $10^{-6}$ torr by continuously running pumping out with a turbomolecular pump. This good level of vacuum allowed us to minimize $LN_2$ consumption rate and to achieve its value of ~ 2 l/h (see section 3.1).

The temperature of the electronics was maintained with an accuracy of ±1° while the ambient temperature in the experimental hall varied from 18°C to 33°C during the data taking period. The recorded temperature inside the rack will be used for correction of the trigger efficiency which is dependent on the individual offsets of the individual PMT signals sent to the discriminators (see section 3.4).

The anode to gate and gate to cathode voltage differences were set to 7.6 and 9 kV, respectively, during the data taking period. The voltage and current of the HV power supply modules (both for PMT and electrodes) were constantly monitored and recorded to the database. The occasional quite rare periods of the HV current instability will be excluded in the further data analysis.

### 4.2 LXe purity

The LXe in the RED-100 detector is continuously purified by circulation through SAES MonoTorr (see chapter 3.2) with a rate of ~ 9 slpm. The electron lifetime evolution during the run is shown in Figure 5. One can see that during the data taking period (mid Jan 2022 – beg Mar 2022) the lifetime value was from ~ 0.5 to ~ 1 ms.

Estimation of error bars of the lifetime measurements was done by the following way. The large sample of waveforms (~100000 events) was recorded by the oscilloscope in an event by event mode. Then, the data were divided into smaller samples with a size used for averaging by the oscilloscope (10000 events). The waveforms in these samples were averaged to obtain ten independent measurements of the lifetime, and the standard deviation of these points from the mean value was obtained. The visible significant drop of the lifetime on Jan 12 from ~ 500 μs to ~ 200 μs was caused by the lengthy replacement job of the circulation pump membrane (during ~ 24 hours).

### 4.3 Background monitoring during data taking

Preliminary measurements of the gamma background in different points of the experimental hall have shown that it is nearly uniform over the hall except of few higher level small areas. The gamma background is approximately by a factor of 5 larger than that measured previously in the MEPhI laboratory. This is due to the reactor building thick concrete walls, ceiling, and floor.

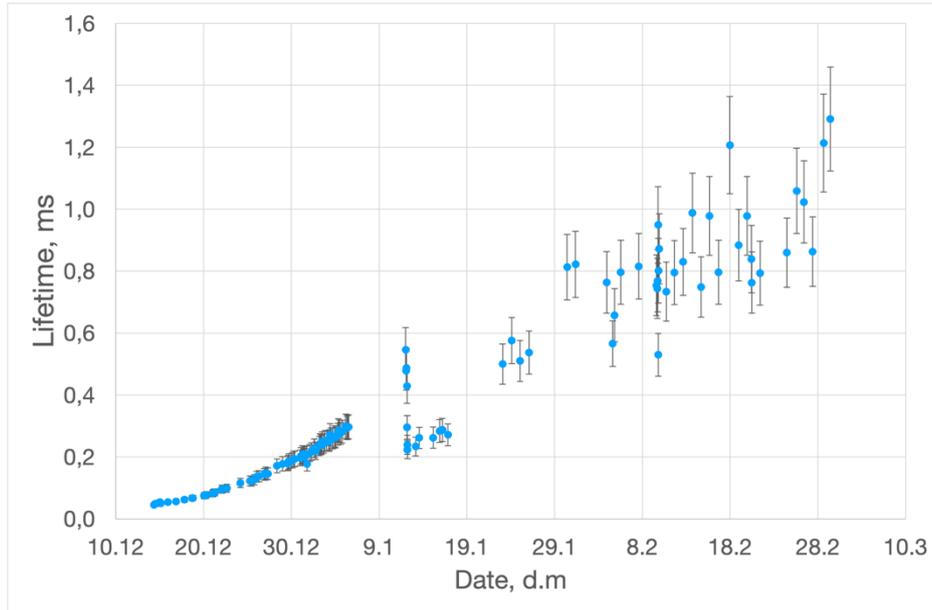

**Figure 5.** Evolution of the electron lifetime during the Dec 2021 – Mar 2022 run.

During the Dec 2021 – Mar 2022 run the NaI[Tl] detector was in the constant location (see section 3.7), and the spectra from the NaI[Tl] detector were acquired within 1200-s live time intervals in an automatic mode. The integrated rate for the energies > 2.2 MeV (the most

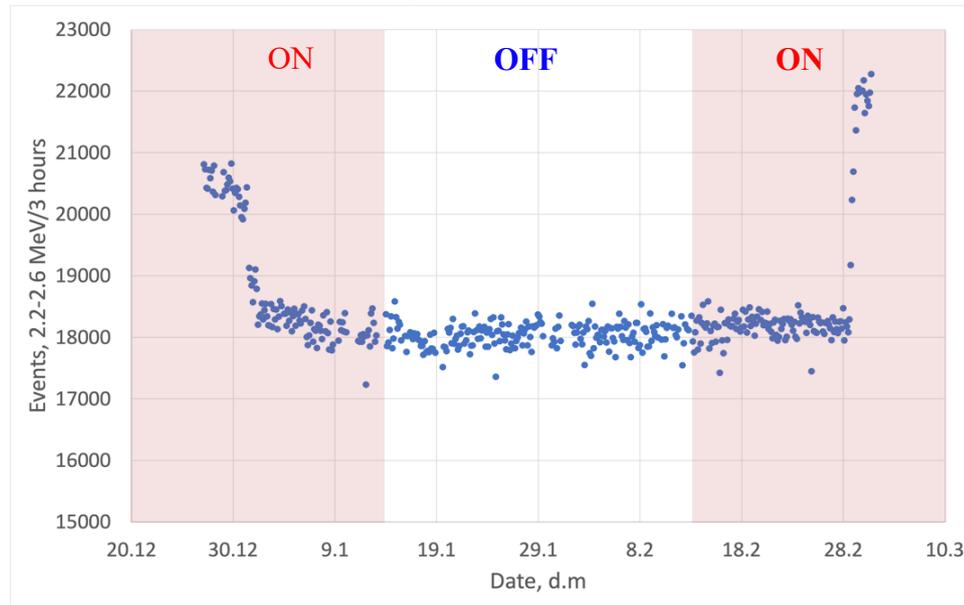

**Figure 6.** Gamma background during the Dec 2021 - Mar 2022 run; the periods of the reactor ON and OFF are shown by color, the vertical dashed lines show the moments of filling (left) and emptying (right) the water tank.

energetic gamma background which can reach the RED-100 detector) is obtained and presented in Figure 6 for 3-h intervals. One can see from this plot that the detector count rate has substantially dropped after the filling of the water tank by water. This is caused by the obscuring

– 11 –

the part of the full gamma flux from the detector by the water tank. The same effect but with the opposite sign one can see after the tank draining. The visible difference between the count rate levels before filling and after draining is explained by the fact that the water tank was partly filled by the moment of start measurements. The presented plot evidences that there is no visible change of the gamma-background during the reactor ON/OFF and data taking period (mid Jan 2022 – beg Mar 2022).

The neutron count rate was monitored during the period of time from the beginning of Feb 2022 to the beginning of Mar 2022. The neutrons were selected within the region on a PSD plot obtained during the calibration measurements (see sec. 3.7). The total count rate of the events in the neutron PSD window vs time is shown in Figure 7. The mean value is ~ 0.23 counts/min.

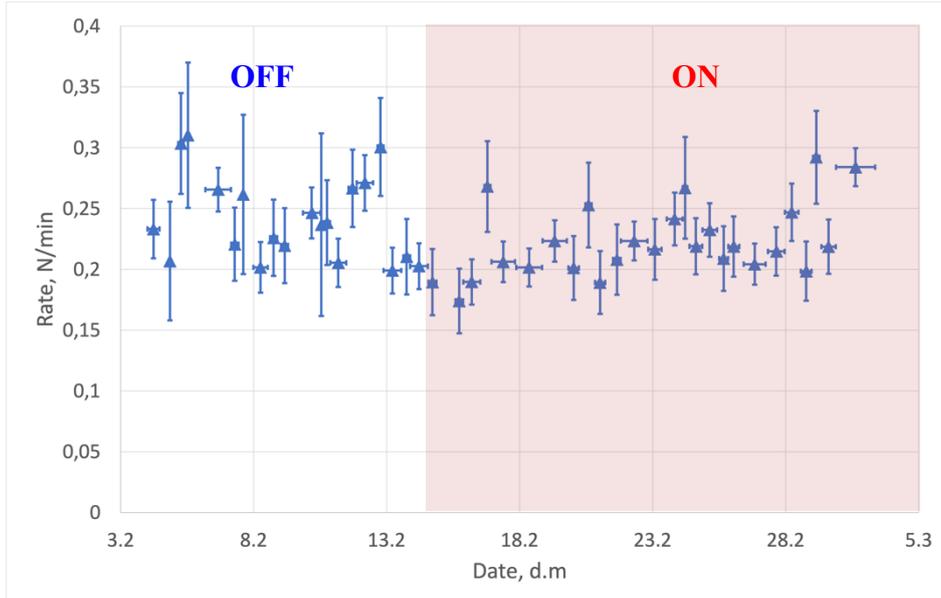

**Figure 7.** Neutron background during beg Feb 2022 – beg Mar 2022; the periods of the reactor ON and OFF are shown by color.

One can see that there is no significant difference in the neutron background count rate during the reactor ON or OFF periods.

The total muon count rate vs time (see Figure 8) is obtained from the muon waveform measurements by the LeCroy digital oscilloscope. It is stable within the accuracy of our measurements.

During the data taking the measured radon level varied from the detection threshold of 30 Bq to ~ 100 Bq with few peaks of up to ~ 300 Bq. No visible enhance of the gamma background was observed at that time.

## 5. Data taking and calibration runs

The data taking started on Jan 14-th and continued till Mar 3-d after finalizing preparation of all RED-100 systems. A data cumulative time plot is shown in Figure 9 (on left) together with a plot of the reactor thermal power. The period of data taking covered both the reactor OFF and ON operation modes except for the period from Jan 25 to Feb 8 when the data was of unacceptable quality because of technical reasons. In the OFF period, the CEvNS data are collected in the same trigger mode for background measurement. The CEvNS-search runs were altered with calibration runs approximately once a week. The total astronomical time of data collection is 8.6 and 12.7 days in the OFF and ON mode, respectively. Percentage of the data acquiring in different trigger modes is shown in Figure 9 (on right). Most of the time (~ 70%) is the science data taking (CEvNS mode).

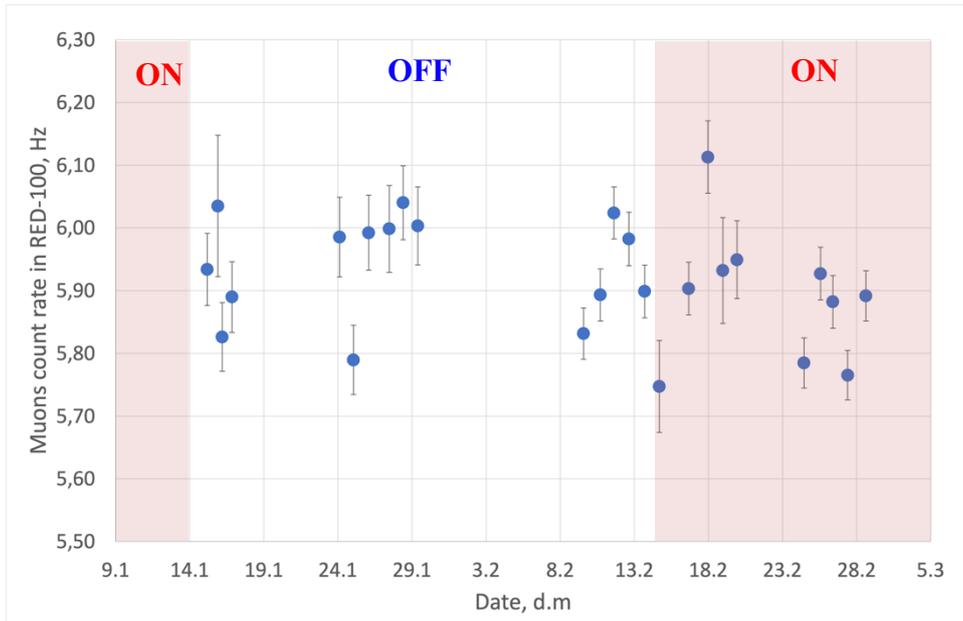

**Figure 8.** Muon background during mid Jan 2022 – beg Mar 2022; the periods of the reactor ON and OFF are shown by color.

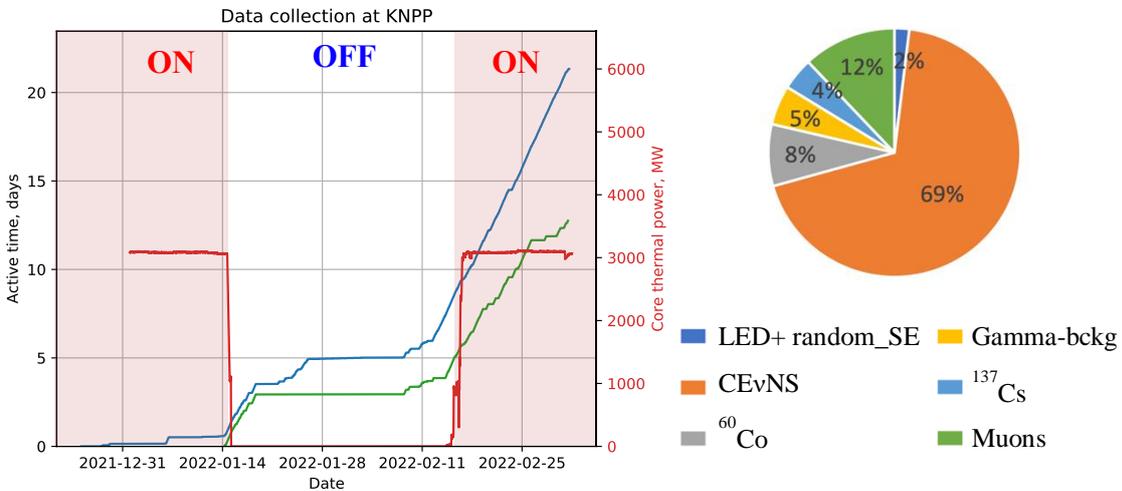

**Figure 9.** On left, the data cumulative time: blue curve for all trigger modes, green curve for CEvNS mode, red curve for the reactor power (right scale), the periods of the reactor ON and OFF are shown by color; on right, the percentage of each mode of data taking time from the total active data taking time.

## 6. Conclusion

For the first time, a massive xenon two-phase emission detector has been deployed at an industrial nuclear power plant in order to detect the coherent elastic neutrino-nucleus scattering process. The experiment to observe this process was carried out with such a massive detector (~ 160 kg in working volume) for the first time. The RED-100 experimental setup was installed at a 19-m distance under the reactor core of the Kalinin Nuclear Power Plant. The brief resume of the RED-100 experiment is the following.

- The feasibility of operation of such a complicated system as a two-phase emission noble gas detector in hard conditions of an operating nuclear power plant was demonstrated for the first time.
- The radiation background conditions are stable and do not depend whether the reactor is ON or OFF.
- The trigger system was tested in the real background conditions. In the test process, the trigger operation mode settings that allow one to select the time intervals with the minimal noise were experimentally tuned.
- The data taking run was carried out in the beginning of the year 2022 covering both the reactor OFF and ON periods, and data analysis is ongoing.
- It has been shown for the first time that a rather massive liquid noble gas detector can be used under the radiation conditions of modern nuclear power plants to detect few-electron signals associated with rare processes.

## Acknowledgments

The authors express their gratitude to the State Atomic Energy Corporation Rosatom (ROSATOM) and International Atomic Energy Agency (IAEA) for the indicated interest to the topic covered by this paper, the Rosenergoatom Joint-Stock Company for administrative support of the RED-100 project, the JSC Science and Innovations (Scientific Division of the ROSATOM) for the financial support under contract No.313/1679-D dated September 16, 2019, the Russian Science Foundation for the financial support under contract No.22-12-00082 dated May 13, 2022, the administrations of: the National Research Nuclear University MEPhI (MEPhI Program Priority 2030), the National Research Center "Kurchatov Institute", the Institute of Nuclear Physics named after G.I. Budker SB RAS, the Tomsk Polytechnic University (Development Program of Tomsk Polytechnic University No. Priority-2030-NIP/EB-004-0000-2022) for support in development of technology of two-phase emission detectors. The authors are grateful to the staff of the Kalinin NPP for their comprehensive assistance in conducting the RED-100 experiment, as well as the scientists of the National Research Center "Kurchatov Institute", who carry out the iDREAM experiment at the Kalinin NPP under supervision of prof. M.D. Skorokhvatov, for assistance in organizing measurements.

## References

[1] A. Bernstein *et al., Colloquium: Neutrino detectors as tools for nuclear security*, *Rev. Mod. Phys.* **92** (2020) 011003.

[2] A. Borovoi and L. Mikaelyan. *The possibility of practical usage of neutrino*, Atomic Energy **44** (1978) 508.

[3] Yu. Klimov et al., *Neutrino method remote measurement of reactor power and power output,* Atomic Energy **76** (1994) 130.

[4] M. Skorokhvatov. *Development of neutrino method for control and monitoring of nuclear reactors, Report at Meeting to Evaluate Potential Applicability of Antineutrino Detection Technologies for Safeguards Purposes*, 17-18 December 2003, International Atomic Energy Agency, Vienna, Austria; www.iaea.org/Publications/Factsheets/English/S1 Safeguards.pdf .

[5] *Final Report: Focused Workshop on Antineutrino Detection for Safeguards Applications,* 28-30 October 2008, IAEA Headquarters, Vienna, 2008.

[6] D.Y. Akimov et al., *Observation of coherent elastic neutrino-nucleus scattering,* Science **357** (2017) no. 6356 1123 [arXiv:1708.01294].

[7] D.Yu. Akimov et al. *Prospects for observation of neutrino-nuclear neutral current coherent scattering with two-phase Xenon emission detector* JINST **8** (2013) P10023 [arXiv:1212.1938]

[8] B.A. Dolgoshein, V.N. Lebedenko, and B.U. Rodionov, *New method of registration of ionizing-particle tracks in condensed matter,* JETP Lett. **11** (1970) 351.

[9] A. Bolozdynya, *Emission Detectors,* World Scientific Publishing Corporation Co. Pte. Ltd.,* 2010. 209 pp.

[10] Akimov D.Yu. et al., *Two-Phase Emission Detectors,* World Scientific Publishing Company, 2021. 250 pp.

[11] A. Bolozdynya et al., Emission detectors. *IEEE Trans. Nucl. Sci.* **42** (1995) 565.

[12] J. Xu, *Status of the CHILLAX detector development,* October 6 – 7 2021 Magnificent CEvNS 2021 on-line Workshop, URL https://indico.cern.ch/event/1075677/contributions/4556726/

[13] H. Bonet et al., *Constraints on elastic neutrino nucleus scattering in the fully coherent regime from the CONUS experiment,* Phys. Rev. Lett. **126** (2021) 041804 [arXiv:2011.00210].

[14] J. Colaresi et al., *First results from a search for coherent elastic neutrino-nucleus scattering (CEvNS) at a reactor site,* Phys. Rev. D **104** (2021) 072003 [arXiv:2108.02880].

[15] G. Agnolet et al., *Background studies for the MINER Coherent Neutrino Scattering reactor experiment,* Nucl. Instrum. Meth. A **853** (2017) 53 [arXiv:1609.02066].

[16] G. Angloher et al., *Exploring CEvNS with NUCLEUS at the Chooz Nuclear Power Plant,* Eur. Phys. J. C **79** 2019 1018 [arXiv:1905.10258].

[17] T. Salagnac et al., *Optimization and performance of the CryoCube detector for the future RICOCHET low-energy neutrino experiment,* [arXiv:2111.12438].

[18] L. J. Flores et al., *Physics reach of a low threshold scintillating argon bubble chamber in coherent elastic neutrino-nucleus scattering reactor experiments,* Phys. Rev. D **103** (2021) L091301 [arXiv:2101.08785]

[19] H. T.-K. Wong, *Taiwan EXperiment On NeutrinO − History, Status and Prospects,* Int. J. Mod. Phys. A **33** (2018) 16 1830014 [arXiv:1608.00306].

[20] V. Belov et al., *The vGeN experiment at the Kalinin Nuclear Power Plant,* 2015 *JINST* **10** P12011.

[21] D.Yu. Akimov et al., *First ground-level laboratory test of the two-phase xenon emission detector RED-100,* 2020 *JINST* **15** P02020 [arXiv:1910.06190].

[22] D.Yu. Akimov et al., *The RED-100 two-phase emission detector, Instrum. Exp. Tech.* **60** (2017) 2 175.

[23] D.Yu. Akimov et al., *Status of the RED-100 experiment,* 2017 *JINST* **12** C06018.

[24] D.Yu. Akimov et al., *A Passive Shield for the RED-100 Neutrino Detector, Instrum. Exp. Tech.* **64** (2021) 2 202.

[25] https://www.cryomech.com/products/lnp120/

[26] D.Yu. Akimov et al., *Purification of liquid xenon with the spark discharge technique for use in two-phase emission detectors, Instrum. Exp. Tech.* **60** (2017) no. 6 782.

[27] D.Yu. Akimov et al., *An Integral Method for Processing Xenon Used as a Working Medium in the RED-100 Two-Phase Emission Detector, Instrum. Exp. Tech.* **62** (2019) no. 4 457.

[28] M.Clausen, *EPICS: Experimental physics and industrial control system, ICFA Beam Dyn. Newslett.* **47** (2008) 56.

[29] https://www.influxdata.com

[30] https://controlsystemstudio.org

[31] https://grafana.com/grafana/

[32] P.P. Naumov et al., *The Digital Trigger System for the RED-100 Detector, Physics of Atomic Nuclei* **78** (2015) 1539.